\documentclass[showpacs,twocolumn,prb]{revtex4}
\usepackage{graphicx}
\usepackage{epsfig}
\usepackage{amssymb}

\def\beq{\begin{equation}}
\def\eeq{\end{equation}}
\def\beqa{\begin{eqnarray}}
\def\eeqa{\end{eqnarray}}

\def\e{\epsilon}

\def\half{{\ss 1\over 2}}

\def\D{\Delta}

\def\e{\varepsilon}
\def\cH{{\mathcal H}}

\def\eff{\mathrm{eff}}
\def\ss{\scriptstyle}

\def\Tr{\mathrm{Tr}}

\def\d{\mathrm{d}}

\def\ss{\scriptstyle}

\def\etal{{\sl et al.}}

\def\nonum{\nonumber \\}

\def\nonum{ \nonumber \\}
\def\eff{\text{eff}}
\renewcommand{\sec}[1]{\vskip 0.5truecm \noindent {\bf #1}. -- }

\begin{document}

\title{Dimer mean-field model for the Ising spin glass}
\author{Yonatan Dubi$^{1,2}$ and Massimiliano Di Ventra$^2$ }
\affiliation{$^{1}$ Theoretical Division, Los Alamos National Laboratory, Los Alamos, NM 87545, USA \\ $^2$
Department of Physics, University of California - San Diego, La Jolla, California 92093-0319, USA}
\begin{abstract}

A dimer mean-field model for the Ising spin-glass is presented. Despite its simplicity it captures some of the essential features of the spin-glass physics. The distribution of the single-spin magnetization is determined from a self-consistent integral equation. By solving the self-consistency condition numerically, we find that there are two temperature scales characterizing the glass transition. At the first, higher temperature, the glass order parameter becomes non-vanishing, and at the second, freezing temperature, it saturates to its maximal value. The effect of magnetic field and the existence of the Almeida-Thouless line are discussed. Finally, it is shown that the information compressibility, defined as the derivative of entropy with respect to energy, diverges at the freezing temperature. This indicates a zero internal temperature and true glassy dynamics with diverging relaxation times.
\end{abstract} \maketitle

The interest in spin-glasses (SGs), by now several decades old \cite{Edwards, Bray, Mezard,Fisher}, stems from the fact that in spite of their seemingly simple description, they display a rich and complex behavior \cite{Stein}. Specifically, quantum SGs have been investigated using elaborate mathematical formalisms such as the replica approach \cite{Mezard} and quantum annealing methods \cite{Brooke}. Studying the exact properties from exact diagonalization of the Hamiltonian requires tremendous computational effort, since the dimension of the Hilbert space increases exponentially with system size \cite{Viet, Young}. Therefore, it has long been recognized that simplified models, which exhibit the glass properties but are easily addressed and solved are highly desirable \cite{Derrida,Gross, Jorg}.

Here, we present such a model based on the short-range Ising spin-glass model. We study its single-spin properties, using a dimer-mean-field approximation for the distribution of the single-spin magnetization. The model is simple and physically transparent, and yet captures some of the essentials of the spin-glass transition. Our main results are as follows. (i) We find that in $D \ge 2$ dimensions there are two important temperature scales. The first, higher temperature, corresponds to the onset of a finite order parameter. The second temperature, is one at which the order parameter saturates to its maximal value. We can thus distinct between the ``glass transition'', at which the order parameter first appears, and the ``freezing transition'', at which it saturates. Interestingly, we find that in 1D there is only a glass transition but no freezing transition. (ii) For $D \ge 2$, there exists a region of magnetic fields in which the freezing transition is supported, thus confirming the existence of an Almeida-Thouless line \cite{Almeida,Stein,Mezard,Young} for this model. (iii) We show that the internal temperature - as experienced by the spins - vanishes at the freezing transition. We argue that consequently the dynamics exhibits a critical slowing down.

{\it Model and method of calculation - }
The starting point of our calculation is the Anderson-Edwards Hamiltonian for the short-range Ising spin glass \cite{Edwards},
\beq \cH=-\sum_{\langle ij \rangle } J_{ij} s_i s_j ~~,\label{hamiltonian1}\eeq where $\langle ij \rangle$ indicates sum over nearest neighbors on a square D-dimensional lattice, $J_{ij}$ are the values $\pm J$ randomly given with equal probability, and $s_i$ is the $z$-direction spin operator for the particle at position $i$.

\begin{figure}[ht]
\vskip 0.5truecm
\includegraphics[width=3truecm]{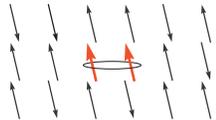}
\caption{(Color online) A schematic of the dimer mean-field model. Out of the whole spin lattice, two neighboring spins are chosen, and their Hamiltonian is written, where the external field experienced by the spins results from the average field generated by their neighboring spins. }\label{model_schematic}
\end{figure}

The next step is to apply a dimer-mean-field approximation. We divide the system into pairs of spins (spin-dimers, see Fig.~\ref{model_schematic}), and we write down explicitly the Hamiltonian of the spin-dimer. In the basis of the dimer states $|\uparrow \uparrow \rangle$,$|\uparrow \downarrow \rangle$,$|\downarrow \uparrow \rangle$,$|\downarrow \downarrow \rangle$, the Hamiltonian is simply given by
\begin{widetext}
\beq
\cH_{\textrm{dimer}}=\left(
  \begin{array}{cccc}
    -J_{12}-h_1-h_2 & 0 & 0 & 0 \\
    0 & J_{12}-h_1+h_2 & 0 & 0 \\
    0 & 0 & J_{12}+h_1-h_2 & 0 \\
    0 & 0 & 0 & -J_{12}+h_1+h_2 \\
  \end{array}
\right) ~~,\label{Hamiltonian2}\eeq
\end{widetext} where $h_1$ and $h_2$ are the magnetic fields experienced by the first and second spins of the dimer, respectively,
and $J_{12}$ is the coupling of the latter. The mean-field approximation is employed by setting  $h_i=h_{\textrm{ext}}-J_i (z-1) \tilde{s}_i$, where $h_{\textrm{ext}}$ is the external magnetic field (in the z-direction), $J_i$ is a (random) exchange coupling, $z$ is the number of nearest neighbors, and $\tilde{s}_i$ is the (average) spin neighboring the spin $s_i$ (see Fig.~\ref{model_schematic}). In fact, for a more exact treatment one should write $h_i=h_{\textrm{ext}}+\sum_{j=1}^{z-1} J_{ij}\tilde{s}_j$, that would amount to accounting for each of the neighboring spins separately. However, we found by comparing numerical calculations on both versions (we used a two-dimensional square lattice for the second) that this does not alter the results.

Since the Hamiltonian is diagonal, it is now a matter of straightforward algebra to evaluate the average spin of either member of the spin-dimer, which is given by (setting $k_B=\hbar=1$)
\beqa
s(h_1,h_2,J_{12})&=&-\frac{1}{Z} \left( e^{-\frac{\text{h1}-\text{h2}-J_{12}}{T}}+e^{-\frac{-\text{h1}+\text{h2}-J_{12}}{T}}+ \right.\nonum &+& \left. e^{-\frac{-\text{h1}-\text{h2}+J_{12}}{T}}-e^{-
   \frac{\text{h1}+\text{h2}+J_{12}}{T}} \right)~~,\label{s1} \eeqa
where $Z=\Tr (e^{-\cH /T})=\sum_{i=1}^{4} e^{\e_i/T}$, with $\e_i$ the diagonal elements of the Hamiltonian, and $T$ is the external temperature.

Since all the $J$s in the above expression are random variables, $s$ and $\tilde{s}$ are also random variables. Therefore, it is the distribution of $s$ which should be determined self-consistently. Since the function $s(J_{12},J_1,J_2,\tilde{s}_1,\tilde{s}_1,h_{\text{ext}})$ is known, the equation for the distribution function of $s$, $g(s)$ reads
\beqa  g(s)&=&\int \d J_{12} \d J_1 \d J_2 \d \tilde{s}_1\d \tilde{s}_2 f(J_{12})f(J_1) f(J_2) \times \nonum & & \times g(\tilde{s}_1)g(\tilde{s}_2) \delta(s-s(J_{12},J_1,J_2,\tilde{s}_1,\tilde{s}_1,h_{\text{ext}})) \label{self-cons-eq}~,\eeqa
where $\delta$ is the Dirac delta-function, and $f(J_{i})=\frac{1}{2}\left( \delta(J_{i}-J)+\delta(J_{i}+J) \right)$ is the distribution function for the couplings.

Eq.~\ref{self-cons-eq} is a central result of this paper and it represents the mean-field self-consistent equation
for this model. In the following, we
solve it numerically for different temperatures, external fields, etc. From the solution of $g(s)$, we can estimate all single-spin quantities, such as the average magnetization $\langle s \rangle$, the so-called Edwards-Anderson glass order parameter, which in this mean-field approximation is simply $\langle s^2 \rangle$ and other properties such as the entropy, energy, etc. Here, we note that naively one could suggest a similar method for a single spin instead of a dimer, yet it turns out that this cannot capture the basic spin-glass properties. The reason is that the glassy behavior is a direct consequence of the competition between disorder and interactions, and thus a dimer structure is the minimal one that includes both.

\begin{figure}[ht]
\vskip 0.5truecm
\includegraphics[width=8truecm]{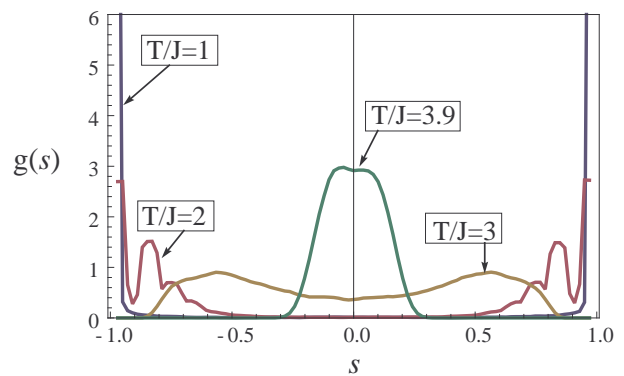}
\caption{(Color online) Distribution function $g(s)$ for a two-dimensional spin glass for temperatures $T/J=1,2,3$ and $3.9$. }\label{dists}
\end{figure}

{\it The glass and freezing transitions - }
We start by calculating the distribution function $g(s)$ for different temperatures $T$ (scaled to units of $J$), with no external field, for a two-dimensional (2D) system. In Fig.~\ref{dists} the distribution function is plotted for temperatures $T/J=1,2,3$ and $3.9$. For $T/J \ge 4$ (which is the number of nearest neighbors, $z$, in 2D) the distribution becomes strongly peaked around $s=0$ (not shown).

As seen from Fig.~\ref{dists}, the distributions are symmetric around $s=0$ for all temperatures (up to numerical errors), and hence $\langle s \rangle_{(T)}=0$. More information about the single spin may be obtained by evaluating the second moment, $\langle s^2 \rangle$, the spin-glass order parameter.

\begin{figure}[ht]
\vskip 0.5truecm
\includegraphics[width=8truecm]{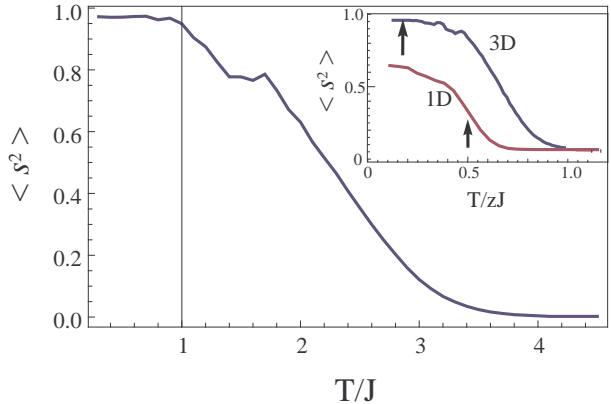}
\caption{(Color online) $\langle s^2 \rangle$ as a function of $T/J$ for the 2D spin glass. At a temperature $T_{\text{G}}=zJ$, $\langle s^2 \rangle$ becomes nonzero, indicating a glass transition. At $T_{\text{f}}=J$, $\langle s^2 \rangle \sim 1$ indicating the suppression of thermal fluctuations and a freezing transition. Inset: the same for 3D and 1D. Arrows indicate the position of $T_{\text{f}}=J$. As seen, in 3D the behavior is similar to that of 2D, but in 1D there is no freezing transition. }\label{s2}
\end{figure}

In Fig.~\ref{s2} we plot $\langle s^2 \rangle$ as a function of $T/J$ for the 2D spin glass. One can observe two relevant temperature scales. The first, $T_{\text{G}}=zJ$, corresponds to the onset of a non-zero $\langle s^2 \rangle$, and we thus recognize this point as the ``glass transition''. However, at $T_{\text{f}}=J$ the order parameter reaches $\langle s^2 \rangle=1$, which implies that thermal fluctuations are completely suppressed. This point thus corresponds to a ``freezing transition''.

In the inset of Fig.~\ref{s2} we plot the same order parameter for 3D and 1D glasses (note that in this mean-field model differences originate only from the number of nearest neighbors $z$). The temperature is now scaled by $T/zJ$, and the arrows indicate the point at which $T/J=1$. In 3D the behavior is similar to that of 2D, but in 1D we find that there is no freezing transition, and $\langle s^2 \rangle$ remains below $\langle s^2 \rangle=1$ down to the lowest temperatures reachable numerically, indicating that the system never ``freezes''. Thus, for this model the critical dimension for the freezing transition is $D=2$.

{\it Effect of magnetic field - }
A standing question in the theory of spin-glasses is the effect of an external magnetic field \cite{Young}. In a seminal paper, Almeida and Thouless predicted that the (mean-field) spin-glass phase is robust against small external fields, and hence the system still has a glass phase and exhibits a glass transition \cite{Almeida}. Large fields, however, polarize the spins regardless of the random interactions, and the glassy behavior vanishes. Thus, there is a crossover from glassy to polarized ground state, the so-called Almeida-Thouless line. It is of interest to see whether our model displays these features, which are supposed to be common to all mean-field SGs.

\begin{figure}[ht]
\vskip 0.5truecm
\includegraphics[width=8truecm]{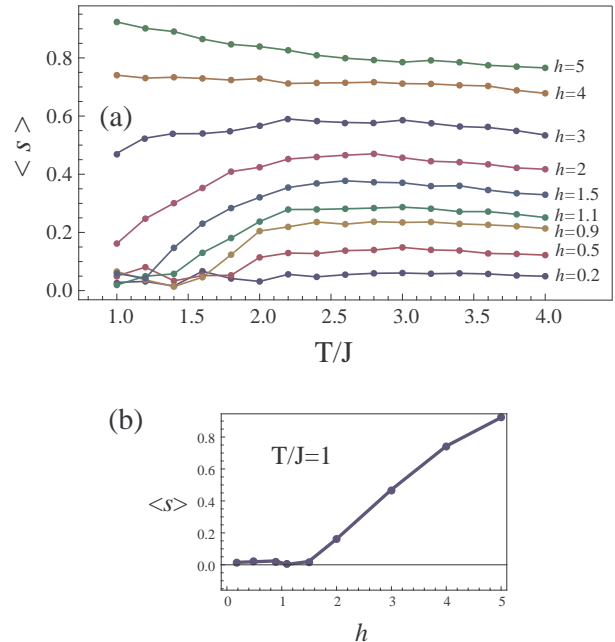}
\caption{(Color online) (a) The single-spin magnetization $\langle s \rangle$ (in two dimensions) as a function of temperature for different values of external magnetic field. (b)  The single-spin magnetization $\langle s \rangle$ at $T/J=1$ as a function of external fields. At small fields the magnetization vanishes, and becomes finite at about $h\sim J$, indicating the existence of an Almeida-Thouless line.  }\label{h-dependence}
\end{figure}

In Fig.~\ref{h-dependence} we show the single-spin magnetization $\langle s \rangle$ in two dimensions, evaluated from the self-consistent solution of Eq.~\ref{self-cons-eq} for finite temperature and magnetic field $h$ (measured in units of $J$). Fig.~\ref{h-dependence}(a) shows $\langle s \rangle$ as a function of temperature for various values of $h$. For small $h$, as the temperature approaches the freezing temperature, the glassy behavior begins to dominate and $\langle s \rangle$ is suppressed. For larger values of $h$, we find finite magnetization even at the freezing transition. In Fig.~\ref{h-dependence}(b) we plot $\langle s \rangle$ at the freezing transition. As seen, at around $h \sim J$ the spins develop finite magnetization, indicating the existence of an Almeida-Thouless line for this model.

{\it Effective temperature and diverging relaxation times - }
In this final section we wish to study the relation between the entropy $S$ and energy $E$ for the present model. As we will show, their relation leads to a vanishing internal temperature at the freezing transition and we will argue that this implies diverging relaxation times, which are one of the hallmarks of glassy behavior.

In a recent paper \cite{Di Ventra}, we have introduced the dynamical quantity $K_I(t)=\frac{dS}{dt'}\left( \frac{dE}{dt'}\right)^{-1} \mid_{t'=t}$ (we set $k_B$=1), which we call the information compressibility. We have shown that for a two-level system open to environments, this quantity defines a dynamic temperature in non-equilibrium processes, by the simple relation $T_{\eff}=\kappa_I^{-1}$ (at equilibrium, $T_{\eff}$ coincides with the thermodynamic temperature), as the temperature of a probe adjusted so that the system's properties are minimally perturbed. When possible, one can use the chain rule and define $\kappa_I$ also for steady-states as $K_I=dS/dE$. This latter quantity is well studied in simple glass models and is known to vanish at the transition \cite{Derrida,Jorg}.

In our model, both the energy and the entropy are easily calculated from the Hamiltonian (Eq.~\ref{Hamiltonian2}). Of course, they are both random variables and depend on the specific realizations of the neighboring spins and the coupling constants. We thus calculate the average values of $S$ and $E$ as a function of temperature (at zero magnetic field), with the use of the distributions obtained earlier (Fig.~\ref{dists}). From these we can construct the function $S(E)$.

\begin{figure}[ht]
\vskip 0.5truecm
\includegraphics[width=6truecm]{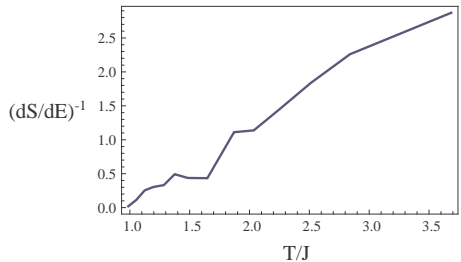}
\caption{The effective temperature, $T_{\eff}=(dS/dE)^{-1}$, as a function of $T$ (in two dimensions). The effective temperature vanishes at the freezing transition, giving an intuitive explanation for the suppression of thermal fluctuations and diverging relaxation time at that point (see text).}\label{dSdE}
\end{figure}

In Fig.~\ref{dSdE} we plot $T_{\eff}=K^{-1}_I=(dS/dE)^{-1}$ as a function of temperature $T$. As can be seen, the effective temperature vanishes at the freezing transition ($T=J$). This provides an intuitive explanation for the suppression of thermal fluctuations: a small change in energy induces a large change in entropy, the number of microstates available to the
system increases, and therefore, the system can explore an ever increasing number of microstates thus reducing the relative fluctuations. This is equivalent to stating that close to the freezing transition fluctuations are dominated by the effective temperature $T_{\eff}$ (which vanishes at the transition) rather than the external temperature.

To understand how a vanishing $T_{\eff}$ leads to a diverging relaxation time (defined as the time it takes the system to restore its equilibrium state after some excitation) we turn to a model introduced by Majumdar \etal \cite{Majumdar} These authors have introduced a general model for glassy dynamics, in which the system is described by an order parameter $\rho$ (in our case, that corresponds to $\langle s^2 \rangle$). They then considered a situation in which transition rates between states of different values of $\rho$ (say $\rho$ and $\rho '$) are determined not by the usual Boltzmann rates, but rather by the following rule:
\beq
W_{\rho \rightarrow \rho'}=\left\{ \begin{array}{c}
                                     e^{-\D S},~~ \D E<0 \\
                                     e^{-\D E /T},~~ \D E>0
                                   \end{array}
  \right. ~~ \label{rates}\eeq
  where $\D E=E(\rho')-E(\rho)$, and $\D S=S(\rho)-S(\rho')$. The authors of Ref.~\onlinecite{Majumdar} have showed that considering such dynamics leads directly to diverging  relaxation times at the transition, since the number of phase-space trajectories that correspond to a decrease in energy decreases as one approaches the glass (or freezing) transition.
  To relate to our finding, we note that for processes which involve a small energy difference, one can write $\D S \sim \frac{dS}{dE} \D E=\D E / T_{\eff}$. This means that, due to the interplay between the change in temperature and change in the number of available microstates, the temperature which defines the relaxation processes - that relax higher energy states into lower ones - essential for equilibrium restoration is $T_{\eff}$ rather then the external temperature.
  As $T_{\eff} \rightarrow 0 $, the rates to go downhill in energy vanish, thus resulting in diverging relaxation times.

  \sec{Summary}
  In summary, we have presented a simple dimer mean-field model to calculate the single-spin properties in the Ising spin-glass. This model, which is physically transparent and computationally cheap, gives insight into the properties of the glass transition. Specifically, we have shown that at the freezing temperature $T_{\text{f}}=J$ thermal fluctuations are suppressed. We have also demonstrated the effect of magnetic field, and showed that the effective temperature of the spins vanishes at the freezing transition. This, in turn, provides an intuitive explanation of the suppression of thermal fluctuations and divergence of timescales at the freezing transition of the Ising spin-glass.

  In the future we plan to extend the model in various directions. One possible direction is to include more than nearest neighbors, and by using real-space renormalization methods we plan to obtain information not only about the single spin but also about correlations between distant spins, which are very important in characterizing the glassy state. Other directions include extending the system to three-dimensional spins (e.g., Heisenberg spin-glasses), including magnetic fields in a general direction, and extending the calculation to larger spins (rather than spin $\half$). These studies can be relevant to various systems such as memory-shape alloys \cite{Kartha,Sherrington}and strained materials \cite{Lookman}.

   We thank F. Krzakala for valuable discussions. This research was funded by DOE under grant DE-FG02-05ER46204.

\end{document}